# Digital Dark Field – Higher Contrast and Greater Specificity Dark Field Imaging using a 4DSTEM Approach

Ian MacLaren[1,*], Andrew T. Fraser[1], Matthew R. Lipsett[1], Colin Ophus[2]

1. School of Physics and Astronomy, University of Glasgow, Glasgow G12 8QQ, UK
2. NCEM, Molecular Foundry, Lawrence Berkeley National Laboratory, Berkeley 94720, USA

*. Ian MacLaren was an Editor of the journal during the review period of the article. To avoid a conflict of interest, they were blinded to the record and another editor processed this manuscript

## Abstract

A new method for dark field imaging is introduced which uses scanned electron diffraction (or 4DSTEM – 4-dimensional scanning transmission electron microscopy) datasets as its input. Instead of working on simple summation of intensity, it works on a sparse representation of the diffraction patterns in terms of a list of their diffraction peaks. This is tested on a thin perovskite film containing structural ordering resulting in additional superlattice spots that reveal details of domain structures, and is shown to give much better selectivity and contrast than conventional virtual dark field imaging. It is also shown to work well in polycrystalline aggregates of CuO nanoparticles. In view of the higher contrast and selectivity, and the complete exclusion of diffuse scattering from the image formation, it is expected to be of significant benefit for characterisation of a wide variety of crystalline materials.

## Introduction

Dark Field Imaging has been part of Transmission Electron Microscopy from early in its development, and of especial importance once the importance of diffraction contrast for understanding images of crystals was realised (Hirsch et al. 1977). As such, it has gone through many iterations of use, with variations such as weak beam dark field, strong beam dark field and so on (Hirsch et al. 1977). Careful use can be used to determine details of dislocations, grain boundaries, planar faults and many more defects (Hirsch et al. 1977). The use with domain or twin structures where additional or split spots appear as a result of the transformation that forms these structures can reveal additional detail (e.g. work by one of the authors on twinned structures in $HfO_2$ thin films (MacLaren et al. 2009)).

In more recent times, *Virtual Dark Field* (VDF) imaging (Rauch and Véron 2014) has been introduced in 4-dimensional Scanning Transmission Electron Microscopy (4DSTEM) / Scanned Electron Nano-Diffraction (SEND) datasets (including scanning precession electron diffraction (SPED)), where an aperture is defined in the diffraction plane (dimensions 2 and 3 in a 4D dataset) and all intensity inside is added up to give a dark field image. Such functionality is supported by a range of open-source codes for data analysis in 4DSTEM (Johnstone et al. 2020, Paterson et al. 2020, Savitzky et al. 2021, Cautaerts et al. 2022), as well as the principal OEM (Original Equipment Manufacturer) software offerings from microscope and microscope peripheral / camera manufacturers. (Details may vary as to what shapes or combinations of apertures are available or easy to define, and as to the efficiency with which the images are computed). Nevertheless, in principle, this does nothing more than the standard TEM implementation. Nor does it change in any way the resolution of the information gained, as the optics that formed the diffraction pattern is the same as in regular TEM and the Abbe criterion still applies – the spatial frequency and frequency range (i.e. diameter) of the diffraction spot used determines the best possible resolution of information that can be determined therewith. It may, however, be noted that the reality may be worse as the electron beam on the sample may be larger than this spatial-frequency limited resolution.

However, it has also long been recognised that the dark field contrast is strongly affected by a range of factors, such as sample thickness and sample tilting (Williams and Carter 2009)). This meant that sample tilting makes dark field imaging of large image areas unsatisfying as the contrast is changing across the whole area (even if the main features can be recognised by the human eye / brain combination despite intensity ramps and so on). A suitable example of this is found in Figure 2 of (MacLaren et al. 2002) in which a thin film is imaged with a dark field images from superlattice reflections. However, whilst this was informative, the contrast change along approximately 1 μm of the film length in cross section is large due to sample bending.

One step forward on addressing this issue is to acquire data with precession electron diffraction in the scan (Vincent and Midgley 1994, Rauch et al. 2010) which reduces the effects of tilt and thickness on spot intensities. A further useful step forward was the idea of Paterson to use regular arrays of spots for zone axis patterns to capture all spots associated with a specific crystal structure or modification thereof (e.g. a set of superlattice spots associated with crystal ordering (McCartan et al. 2021)). The aperture positions can themselves be determined by detecting spots, sorting into a 2D lattice with two defined lattice vectors, and then using this to define an array of apertures within the pixel range of the detector (Paterson et al. 2020). Such an approach has recently been used by (Shao et al. 2023) for the characterisation of domain structure in $MoS_2$ sheets.

More recently, a different approach for 4DSTEM processing has been introduced where instead of adding up intensity in image areas, spots are detected and stored as lists of their key parameters, a.k.a. *points lists* (containing information such as position in x and y in the diffraction pattern, intensity, calibrated positions after determination of pattern centre and addition of a calibration from pixels to reciprocal length, and even indices, if indexed to a particular crystal structure) (Savitzky et al. 2021). Obviously, use of lists is a much more *sparse* representation of a pattern, requiring orders of magnitude less in storage space and simplifying processing. Initially, points lists have mainly been used in strain analysis from zone axis patterns(MacLaren et al. 2021) and for Automated Crystal Orientation Mapping (ACOM) (Ophus et al. 2022). To some extent, this evolution in 4DSTEM is merely catching up with what the crystallography community has been doing for many years, where the input to crystal structure refinements is typically lists of hkl indices and diffracted intensities for a given set of cell parameters, e.g. (Zandbergen et al. 1997, Jansen et al. 1998). The present paper shows that such lists can also be used for a more effective dark field imaging approach using 4DSTEM and compares this to standard Virtual Dark Field approaches for imaging using the same dataset.

## Experimental Details

A $La_2CoMnO_6$ thin film was grown on $SrTiO_3$ as described previously by(Kleibeuker et al. 2017). This was prepared for transmission electron microscopy by a standard FIB lift-out method, as also described therein (Kleibeuker et al. 2017). Scanning precession electron diffraction was then performed on a suitable area of the thin film and substrate using a NanoMEGAS TopSpin system using a MerlinEM detector and readout system (MacLaren et al. 2020, McCartan et al. 2021) to record the data in an electron counting mode with minimal readout noise. The microscope used was a JEOL ARM200F operated at 200 kV in standard TEM-L diffraction mode with a camera length of 80 cm, a spot size 5 (the smallest), and a 10 μm condenser aperture (the smallest), giving a convergence angle of ~1.3 mrad and a probe diameter of about 2.3 nm. A SPED dataset was recorded with a step size of 2 nm.

A sample of CuO nanoparticles was dispersed in propan-2-ol and dropped onto a lacey carbon film grid. SPED data was recorded as above, but with a step size of 5 nm.

**The Digital Dark Field Method and a Comparison to Previous Dark Field Approaches**

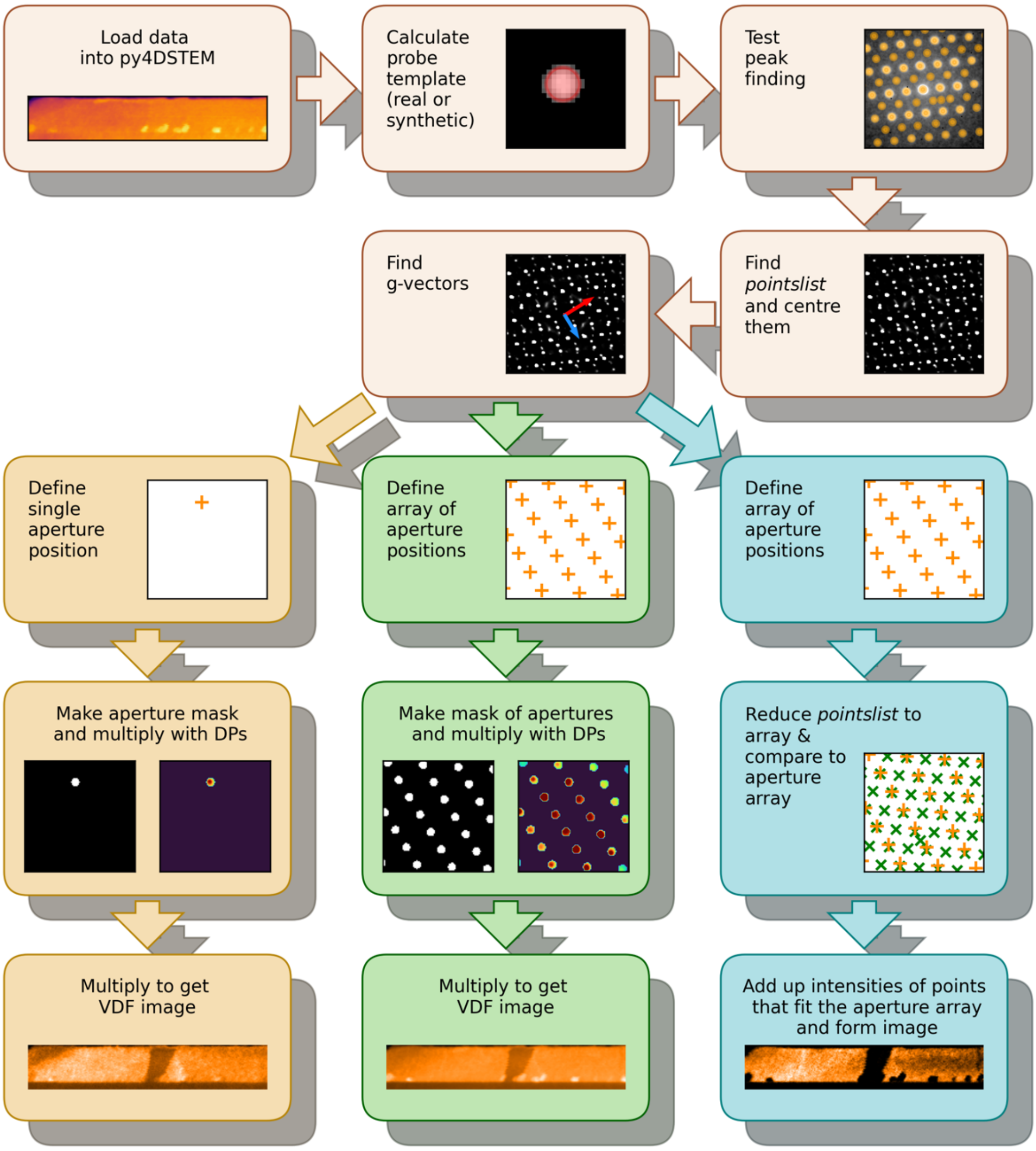

*Figure 1: Diagrammatic workflow of different methods to make a dark field image from a scanned electron diffraction dataset with distinct, non-overlapping, diffraction discs. The most frequently used conventional approach is the one on the left. Some recent studies have used the central path. The new Digital Dark Field approach is the one on the right.*

Figure 1 shows a diagrammatic workflow, incorporating some real images from a dataset showing different dark field calculation approaches for 4DSTEM / SEND / SPED data – in each case, images are shown for operations on a single diffraction pattern representing the whole 4DSTEM dataset. In all cases, the dataset is stored on disk in hdf5 format and then loaded into py4DSTEM. A representative probe is determined from an area of the dataset and a peaks list is determined for every diffraction peak in every pixel of the dataset using methods in py4dstem already discussed previously (Ophus et al. 2022); examples may be found at https://github.com/py4dstem/py4DSTEM_tutorials (more details of the probe template and peak finding in Supplemental Materials). The pattern centre is found for each diffraction pattern and a centred peaks list is used in all succeeding operations (Savitzky et al. 2021, Ophus et al. 2022). In what is basically a single crystal dataset like that from the epitaxial thin film used as the main example in this paper, it is easy to then determine two g-vectors, $\mathbf{g}_1$ and $\mathbf{g}_2$ representing the regular lattice, which dominates the dataset, using tools from the strain analysis module. This is the point at which methods diverge.

It is possible to do just single aperture dark field using one aperture position defined by some suitable multiples of $\mathbf{g}_1$ and $\mathbf{g}_2$. This can then be used along with a suitable aperture radius parameter to then calculate a mask which is simply multiplied through the dataset and the result summed to generate a dark field image, which is shown in pale yellow boxes on the left.

It is possible to do multiple aperture dark field by using $\mathbf{g}_1$ and $\mathbf{g}_2$ to produce an array of aperture positions, possibly with some offset from having the pattern centre as the origin (especially useful for superlattice spots), and possibly only within a given radius or range of kx and ky values (in raw pixels – there is no need for calibration here, especially as uncalibrated array indices are needed for addressing locations in the arrays). Once an array has been generated and converted to a mask, this can simply be multiplied through the dataset and the result summed to generate a dark field image as before. This is shown in the pale green boxes in the centre and was the approach used in McCartan *et al.*(McCartan et al. 2021).

It is also possible to generate the same array of aperture positions, but instead of creating a digital mask, to simply compare lists and this imaging method is shown in blue boxes to the right. Whilst this could be done by iterating through lists and using logic statements, that is inefficient and slow. Far faster is to:

- convert the *points list* to a *N* row × 5 column array, where the 5 columns are $Q_x$, $Q_y$, $I$, $R_x$, $R_y$, and $N$ is the number of *points* (= detected diffraction spots) in the whole dataset
    - $Q_x$, and $Q_y$ are the reciprocal space positions of the diffraction spots along axes 0 and 1 of a diffraction pattern [or axes 2 and 3 of a 4D data array], which are vertical-down and horizontal-right in python, $I$ is the integrated intensity of a diffraction spot, $R_x$ and $R_y$ are the real space positions of the scan for the diffraction pattern containing that diffraction spot and are measured along axes 0 and 1 of the scan [also axes 0 and 1 of the 4D data array], and these are vertical-down and horizontal-right, as before

- calculate differences in position in the diffraction plane $\delta_{Qx}$ and $\delta_{Qy}$ for each entry in the array of diffraction spot positions from each of the positions in the aperture array
  - $\delta_{Qx} = Qx_{spot} - Qx_{aperture}$
  - $\delta_{Qy} = Qy_{spot} - Qy_{aperture}$
- calculate the shortest differences $\delta_{Qr}$ using Pythagoras for each of these entries
  - $\delta_{Qr} = \sqrt{\delta_{Qx}^2 + \delta_{Qy}^2}$
- and discard all entries in the array for which $\delta_{Qr} > tol$, some tolerance value, typically of the order of 1 pixel (tolerances of 1-3 pixels have been used in all applications of code in our group so far, with spot radii of 3-5 pixels and typical smallest spot spacings in the principal lattice of 20-20 pixels (note subpixel precision measurement of feature centroids is a common feature in electron microscopy, such as atomic-resolution imaging and disk position measurements over many years). If some strain or crystal rotation is present, and the magnification is lower, then a larger tolerance may be needed. If it is a higher magnification, then little by way of effects of strain or tilt are likely on spot position and a smaller tolerance may be possible. Setting by trial and error is advised – large enough to include all areas of interest related to the spot array wanted, but small enough to exclude any other closely placed spots from something else).
- add the intensities for each *point* for each $R_x, R_y$ position into the relevant coordinates in a new image. For $n$ *points* matching aperture positions for the probe position $R_x, R_y$, the intensity for that position is evaluated as:
  - $I_{Rx,Ry} = \sum_{i=0}^{n-1} I_{Rx,Ry_i}$

  this therefore only adds up definite diffraction spots sitting on the defined lattice.

Figure 2 shows the creation of dark field images from the dataset, indicating in each case the spot(s) used to create each image superimposed on a diffraction pattern average from the box shown. Figure 2a) shows an annular dark field image, where both the film and substrate are relatively bright, with some dark area above (vacuum or surface carbon), and a few brighter particles at the substrate-film interface (previously shown to be CoO (Kleibeuker et al. 2017). The remainder of the images are all made from weak superlattice spots that should be specific to the ordering of the LCMO, see (Woodward and Reaney 2005) for a fuller discussion of superlattice spots and tilt systems in perovskites. Figure 2b) shows a conventional virtual dark field image made from one diffraction spot close to the pattern centre, the diffraction spot is indicated with an orange disc (a shift of [1/2,-3/2] of the two base lattice vectors) in the pattern of Figure 2j). This is a superlattice spot of this ordered perovskite structure, and is weak, but clearly present. This is not dissimilar to the 4DSTEM imaging recently performed by (Meza et al. 2023) using superlattice spots in $PbSr_2S_3$. It is clear that there is a significant background intensity in every part of the image, including in the substrate below and in the platinum film above. Additionally, there is significant intensity variation along the length of the film in this image, since there is sample bending and the diffraction condition is changing a little with position. The diffraction pattern of Figure 2j is the average diffraction pattern for the box in Figure 2b).

Figure 2d shows another virtual dark field image made using a different spot close to the pattern centre ([1/2,-1/2]), indicated by a blue-green disk in the diffraction pattern of Figure 2j). In this case, additional areas appear bright at the base of the film and close to the interface with the substrate. A diffraction pattern from one of these areas (as indicated by the box in Figure 2d) is shown in Figure 2k), which immediately explains the bright contrast. In this case, these crystals are something totally different to the intended target of the dark field imaging, and the diffraction spots are not in exactly the same place, but still contribute intensity within the aperture resulting in these particles also appearing in this dark field image. These are believed from prior work to be rock-salt structured CoO (Kleibeuker et al. 2017).

Using the approach of Paterson *et al*. (adapted for use in py4DSTEM), we can instead image with multiple apertures covering the $[(n/2, m/2]$ spots, where $n$ and $m$ are odd integers. This is the whole family of spots that include those used to make 2b) and 2d). The resulting VDF image is shown in Figure 2f). Figure 2l) is the average diffraction pattern corresponding to the box on Figure 2f), and the purple overlays mark the array of apertures. Contrast is more even than in the single aperture Virtual Dark Field images, but the CoO particles still appear bright.

Finally, Figure 2h) was made using a different array of apertures ($[n, m/2]$ spots, where $n$ is an integer and $m$ is an odd integer). This shows a different area of the film brighter than in the other images, and the corresponding diffraction pattern averaged from the box on Figure 2h) is shown in Figure 2m) with aperture overlays. This is clearly a different domain with a different crystallographic orientation.

In all cases, the images were made with apertures of the same radius as the probe radius, as determined before finding the peaks in the image. What is clear in all these conventional Virtual Dark Field images is that when making them with weak superlattice spots, there is a lot of background intensity in places that have nothing to do with that set of superlattice spots (e.g. the simple perovskite $SrTiO_3$ substrate). The reason for this is that diffuse scattering (both inelastic and pseudo-elastic [i.e. Thermal Diffuse

Scattering]) is ever-present and still contributes everywhere there is a crystal with significant scattering.

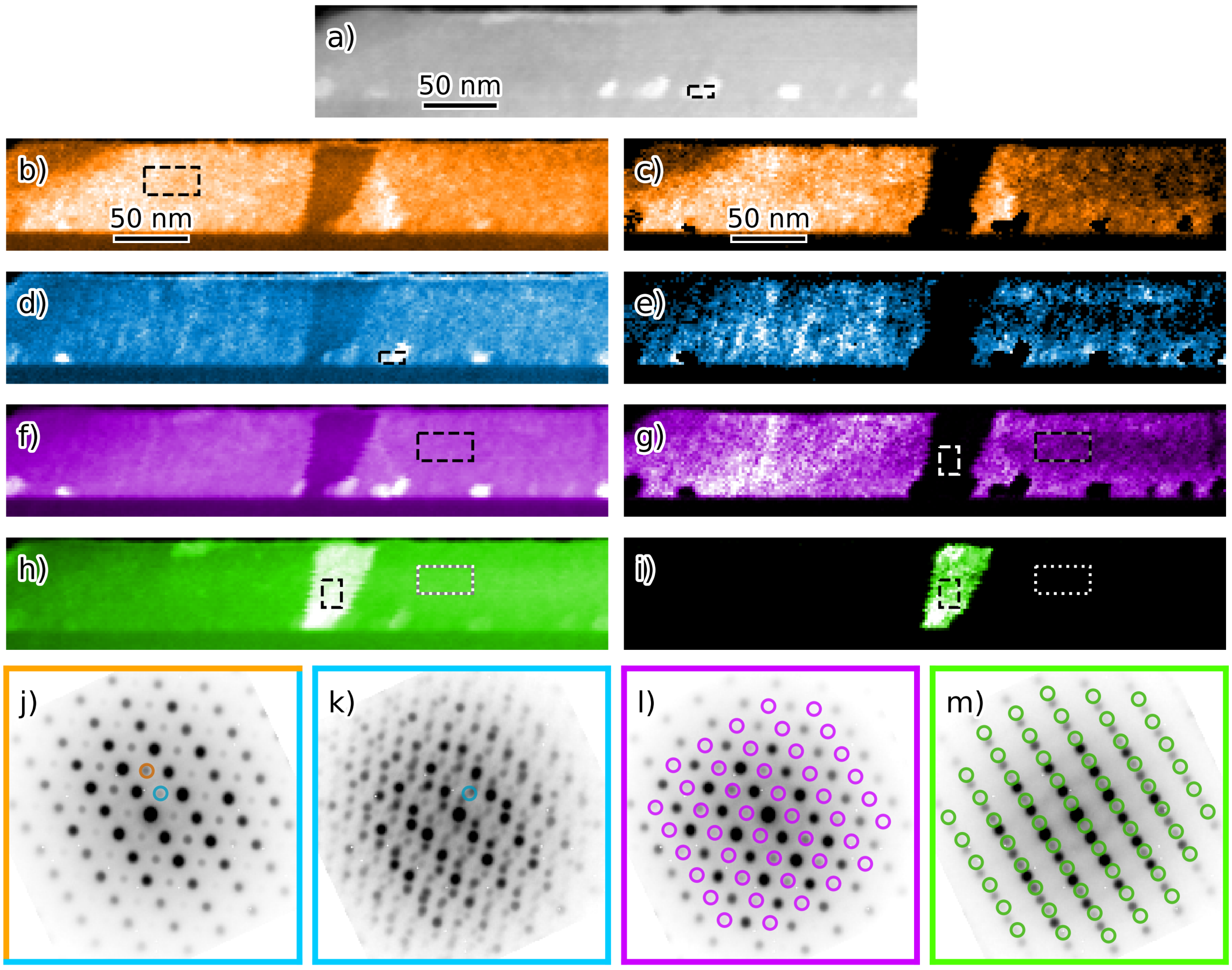

*Figure 2: Comparison of different methods for making virtual dark field images from the same dataset showing both images and some representative diffraction patterns: a) annular dark field image; b-i) are comparisons of the conventional aperture summation (left column) and the new Digital Dark Field approach (right column). b) and c) are created with the orange-ringed spot in j); d) and e) are created with the blue-green-ringed spot in j) and k); f) and g) are created from the purple-ringed lattice of spots in l); and h) and i) are made using the green-ringed lattice of spots in m). . j) is summed from the box shown in b); k) from the box shown in d), l) from the box shown in f; and m) from the box shown in h). All diffraction patterns had a power of 0.2 applied to the values for display purposes only, so that the weak superlattice reflections are more clearly visible in the figure.*

The comparison of the Digital Dark Field method is shown on the right side of the figure using exactly the same single aperture positions or arrays of aperture positions to calculate the images. Figures 2c) and 2e) are the counterparts to the single aperture VDF images of 2b) and 2d). Whilst these are a little noisy (and using just a single dark field spot is not recommended as the best method), they already show clear differences to the conventional VDF. This is especially clear at the interface, where the DDF images have large gaps everywhere where a CoO particle is present (whereas the conventional VDF usually shows these at similar or brighter contrast than the perovskite). This is much clearer in the multiple diffraction spot Digital Dark Field image of 2g), made with the whole $[n/2, m/2]$ array of aperture/spot positions. Now all areas of perovskite film with this ordering are bright, and there is almost no intensity from the CoO, or from the central region of the film with a differently ordered domain, or from the substrate. Figure 2i) shows another multiple diffraction spot Digital Dark Field image using the $[n, m/2]$ spot array. This has dramatically higher contrast than the conventional VDF of 2h). Just to put this on a quantitative basis, and using the boxes indicated on Figs 2f-i) for the calculation areas and calculation using $(I_1 - I_2)/I_2$ using the mean intensity in the boxes. $I_2$ in each calculation is measured in the black dashed box and $I_1$ in the white dotted box on any given image. The contrast levels are compared in Table 1 below.

| Chosen features | VDF Contrast | DDF Contrast |
|---|---|---|
| Main domain – 2f)/2g) | *0.42* | *520* |
| Centre domain 2h)/2i) | *0.47* | ∞ |

*Table 1: Comparison of contrast between VDF and DDF using the boxes marked in Figure 2.*

The clear conclusion is that contrast in the Digital Dark field images is much higher than in the conventional Virtual Dark field. This happens because the intensity quickly drops to near zero in DDF where the spots being summed for intensity disappear (not quite zero in one case, as noise in the data does give a small number of “spots” detected in the dark area for the spot detection settings chosen (see Supplemental Materials), which are a trade-off between detecting all real spots and counting some noisy pixel clusters as real). Other intensity in the area of the spot array in DDF has no impact on the intensity count. This includes diffuse inelastic scattering, diffuse streaks (e.g. from disordering or from phonons in materials with anisotropic phonon modes, such as diamond-structured semiconductors(Cowley 1995)), or nearby diffraction spots from other phases. The latter point is why it is so much better at discriminating between the perovskite film and rock-salt impurity. In other words, this DDF imaging demonstrates *greater specificity* than regular VDF imaging.

There are, of course, weaknesses in this approach, despite the advantages over regular virtual dark field imaging that are quantified above. Principally, all these come down to the dependence on constructing a complete set of diffraction spot positions, a *points list* for the dataset. One issue with that is that this is currently done in post-processing on complete datasets and is therefore intrinsically not live. Some implementations of VDF give live imaging of the results as a dataset is being collected e.g. (Clausen et al. 2020). It is, however, not that live imaging with DDF is impossible, but rather that it would need to be coded as a later extension, whereby points for each diffraction pattern are determined immediately after acquisition and then the DDF result for that pixel calculated. Secondly, and this is certainly an issue with this dataset, it only works if you detect the full array of points. This is much more of an issue for weak superlattice spots that can be partially affected by the background intensity from neighbouring brighter spots. In short, if you don't get all the spots, or are marginal at doing so, intensity counts may vary rather from one pixel to another. So, choosing a good template of a diffraction spot for cross-correlation that reasonably well represents the weak spots is a challenge. In our experience, choosing a bright spot from vacuum may be a poor match, and using synthetic probes (i.e. calculated ones of same radius as original diffraction discs) is recommended. Even with quite some care about this issue, Figs 2c) and 2e) are visibly noisy, especially on the right as the sample gets thicker and more plural scattering is present, making weak spot detection harder. Moreover, some areas which appear dark on the right of Figure 2e) have no discernible [1/2,-1/2] diffraction spot when the diffraction pattern is examined by eye, so any intensity in the conventional VDF image for those pixels is principally diffuse scattering. And this brings the third point, that the intensity in Figure 2g) does decline a little to the right in the centre of the film. It does seem that this technique in separating the intensity in the sharp spots from the diffuse background really does see some slight increase in orientation contrast (aka bend contours) and thickness effects compared to multiple aperture VDF. This may be expected, as we are explicitly just looking at the coherent diffraction and ignoring anything that was scattered in a diffuse way, even fairly close to the diffraction spots.

Nevertheless, even with the above caveats applied, this is an extremely useful imaging technique that may find wide application because of the much-improved selectivity and contrast of the images in complex systems with closely spaced diffraction spots. Furthermore, it is straightforward to extend this approach beyond epitaxial films to polycrystalline materials, simply using different ways of setting up the array of aperture positions, for instance as a single row of apertures with just one **g**-vector for a 2-beam condition. A demonstration of this for some CuO nanoparticles is shown in Figure 3, which shows an ADF image (23-42mrad) of a cluster of nanoparticles with little discernible internal structure. Six different positions were chosen in the dataset (and indicated on the ADF image) from which to extract diffraction patterns that could be used as templates for an array of apertures for forming a dark field image. (Note, this is not a formal segmentation (Bergh et al. 2020) or decomposition of the dataset, merely a demonstration of some of the features of the dataset). Both classic VDF images and DDF images were prepared using arrays of aperture positions chosen to fit the six different diffraction patterns – 3 each of 2D arrays of spots (for particles near a zone axis) and 1D arrays of spots (for particles close to a 2-beam condition). In both cases, the six dark field images were combined into single colour images by:

- Creating a colour image from each using the $HLS$ (Hue-Lightness-Saturation) model where:
  - $H = n/7$ (where $n$ is the number of the image, $H$ ranges from 0-1, 0 and 1 being red, 0.33 being green, 0.67 blue and so on)
  - $L = \frac{0.9I}{I_{max}}$
  - $S = \frac{I}{I_{max}}$

  which produces a color wheel like colour – intensity map
- Combining the 6 maps using a *lighten* type algorithm, where for each pixel and each of the RGB colour channels, the brightest value (closest to one) is chosen as the final brightness (very similar to implementations in popular image-processing packages)

As before, the specificity is better with DDF and the background field is basically black, whereas there is definitely some diffuse intensity across the agglomerate area with VDF (and sometimes additional crystals seen in VDF which must have some similar diffraction spot positions to those in the target crystal).

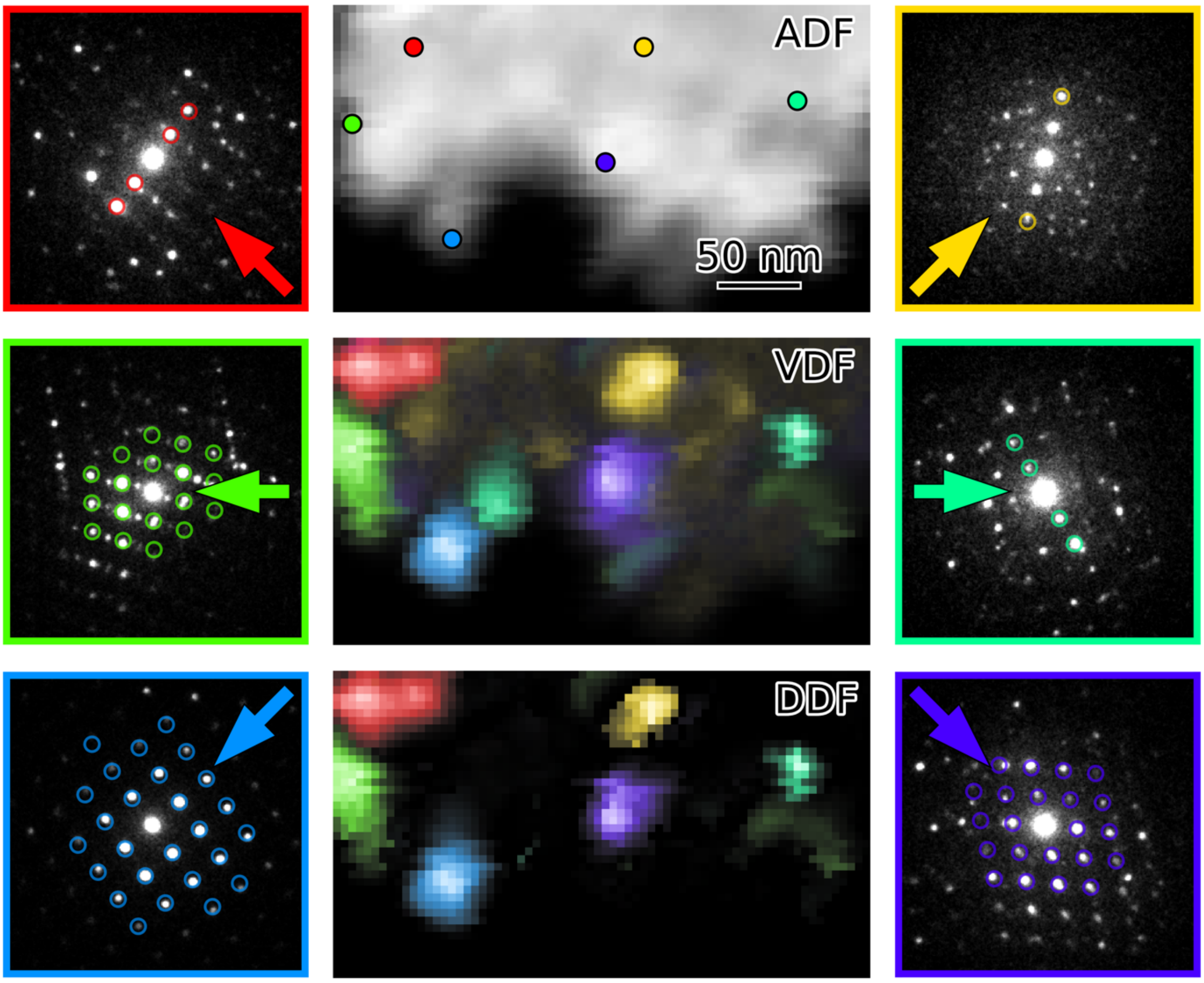

*Figure 3: A comparison of multiaperture VDF and DDF imaging applied to an agglomerate of CuO nanoparticles. Diffraction patterns from 6 different points in the dataset are used as the prototypes for detecting a suitable array of spots for forming a dark field image. Each spot position is marked with a coloured dot on the ADF image (calculated for scattering angles of 21-42mrad, top). Arrows and coloured frames matching the dot colours and relative positions in the scan indicate the 6 diffraction patterns. Classic VDF and DDF images are calculated using the shown arrays of spots and combined into single, multicolour images using each diffraction pattern colour to colour the respective dark field image, and combining the six colour images for each case using a "lighten" algorithm.*

An alternate VDF imaging method was recently used to determine in-plane orientations of the normals to edge-on nanocrystallites using conventional VDF imaging with apertures set by sector around a diffraction ring for a polycrystalline material (Wu et al. 2022). That could also easily be transformed to the DDF paradigm using disc detection (provided discernible diffraction discs are produced), the calculation of the azimuthal angle for the diffraction spot of the correct radius, and subsequent plotting on a colour-wheel.

**Conclusion**

A new method for dark field imaging has been developed using scanned electron diffraction datasets (in this case recorded with precession) which instead of working by simple summation of intensity in areas of the back focal plane, works on lists of detected diffraction peaks. This builds on previous advances in software for 4DSTEM where such sparse representations of diffraction patterns were already in use for other purposes such as strain or orientation/phase mapping. The resulting technique has much higher specificity for distinguishing phases or crystals with close lying diffraction points and completely excludes diffuse scattered intensity in the diffraction patterns. As such, it results in vastly improved contrast in images and much greater certainty that the image only represents the crystal or domain of interest. It is anticipated that this will be of significant use for characterisation of crystalline materials in the electron microscope.

**Code and Data Availability**

The code and raw data used to perform this work will be archived for reader download at https://doi.org/10.5525/gla.researchdata.1771. The code functions are part of py4DSTEM (py4dstem/py4DSTEM/process/diffraction/digital_dark_field.py) together with inline documentation in the github library at https://github.com/py4dstem from 0.14.17 onwards. An provision of an example notebook will be provided in https://github.com/py4dstem/py4DSTEM_tutorials/tree/main/notebooks in due course.

**Acknowledgements**

Dr Josee Kleibeuker and Prof Judith L. Driscoll of the University of Cambridge are gratefully acknowledged for the provision of the $La_2CoMnO_6$ thin film sample.  The CuO sample was provided by Prof. Vernon Phoenix of the University of Strathclyde. NanoMEGAS and Quantum Detectors are gratefully acknowledged for their help in developing the system that allowed scanning precession electron diffraction to be recorded to an electron-counting hybrid pixel sensor, together with support from the EPSRC Impact Acceleration Account (EP/ R511705/1).

# Digital Dark Field – Higher Contrast and Greater Specificity Dark Field Imaging using a 4DSTEM Approach – Supplemental Materials

Ian MacLaren[1,*], Andrew T. Fraser[1], Matthew R. Lipsett[1], Colin Ophus[2]

1. School of Physics and Astronomy, University of Glasgow, Glasgow G12 8QQ, UK
2. NCEM, Molecular Foundry, Lawrence Berkeley National Laboratory, Berkeley 94720, USA

## Spot Detection Details

Figure S1 shows the probe used for peak finding via cross-correlation. This synthetic probe, based on an experimental one, (but with circular symmetry) gave the best results in finding the largest number of real peaks versus noise peaks that we could find.

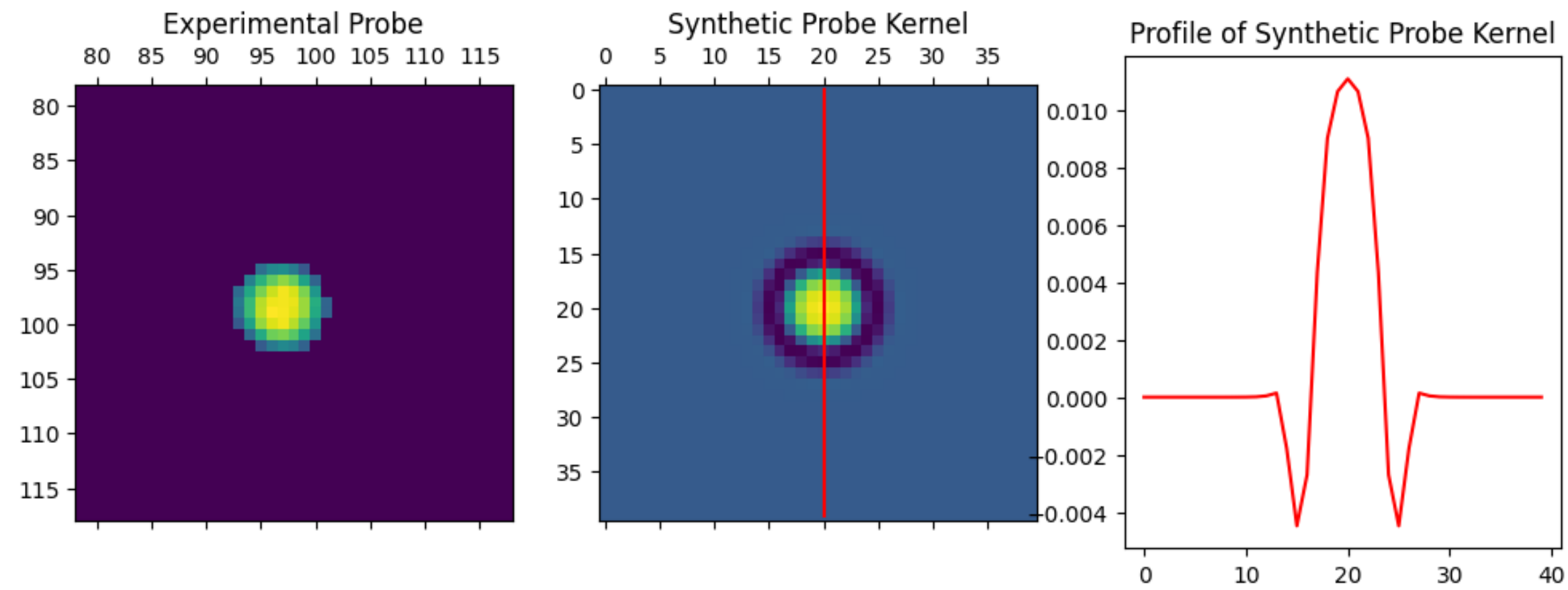

*Figure S1: The probe used for peak finding via cross-correlation: a) an average primary beam from a box 10 pixels deep and 50 wide with top left corner at 10, 200 in the scan dimensions; b) a synthetic Kernel created from this with same semiangle and an edge boundary of 0.8xsemiangle; c) a vertical profile through this kernel.*

Figure S2 shows a synthetic ADF image of the scan area with six colour-coded dots on showing places where example diffraction patterns were extracted for testing of the peak finding settings. Figure S3 then shows the diffraction pattern at every scan point with the detected peaks overlaid as coloured circles. Some, such as the cyan-framed one (top right) and deep pink one (bottom right) mainly found main perovskite diffraction spots, as well as face-centre superlattice spots with few false detections. The orange one (bottom centre) had a few more peaks in odd places, some of which might be real and some of which are probably just noise. The magenta one (top centre) has a lot of extra spots from the CoO inclusion, but these are real, and well-detected. Finally, the most difficult case was the red one (top left) from the area where the superlattice spots were at half spacings of one of the principal reciprocal lattice vectors and rather close to the main points. To detect as many of these as possible, the minimum intensity per spot had to be set quite low, but this led to a few clusters of intensity in the noise also being reported as spots. Ultimately, such false detection of noise as spots hardly matters to the technique as the intensity in the noise spots is very low, and also not on any of the expected aperture grids, so this makes little or no difference to DDF intensity in an image.

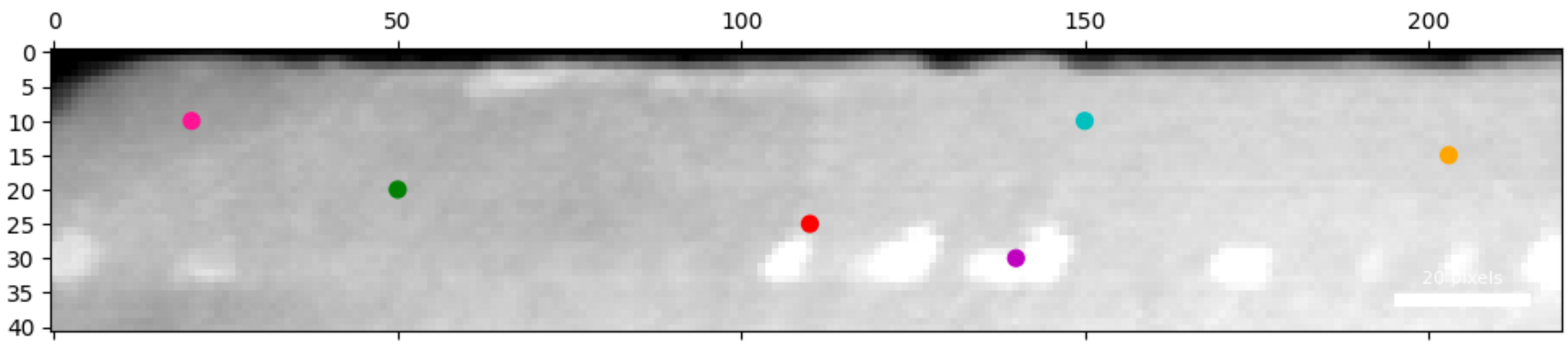

*Figure S2: A synthetic ADF image of the area with six scan points selected from which diffraction patterns were extracted to test the peak finding.*

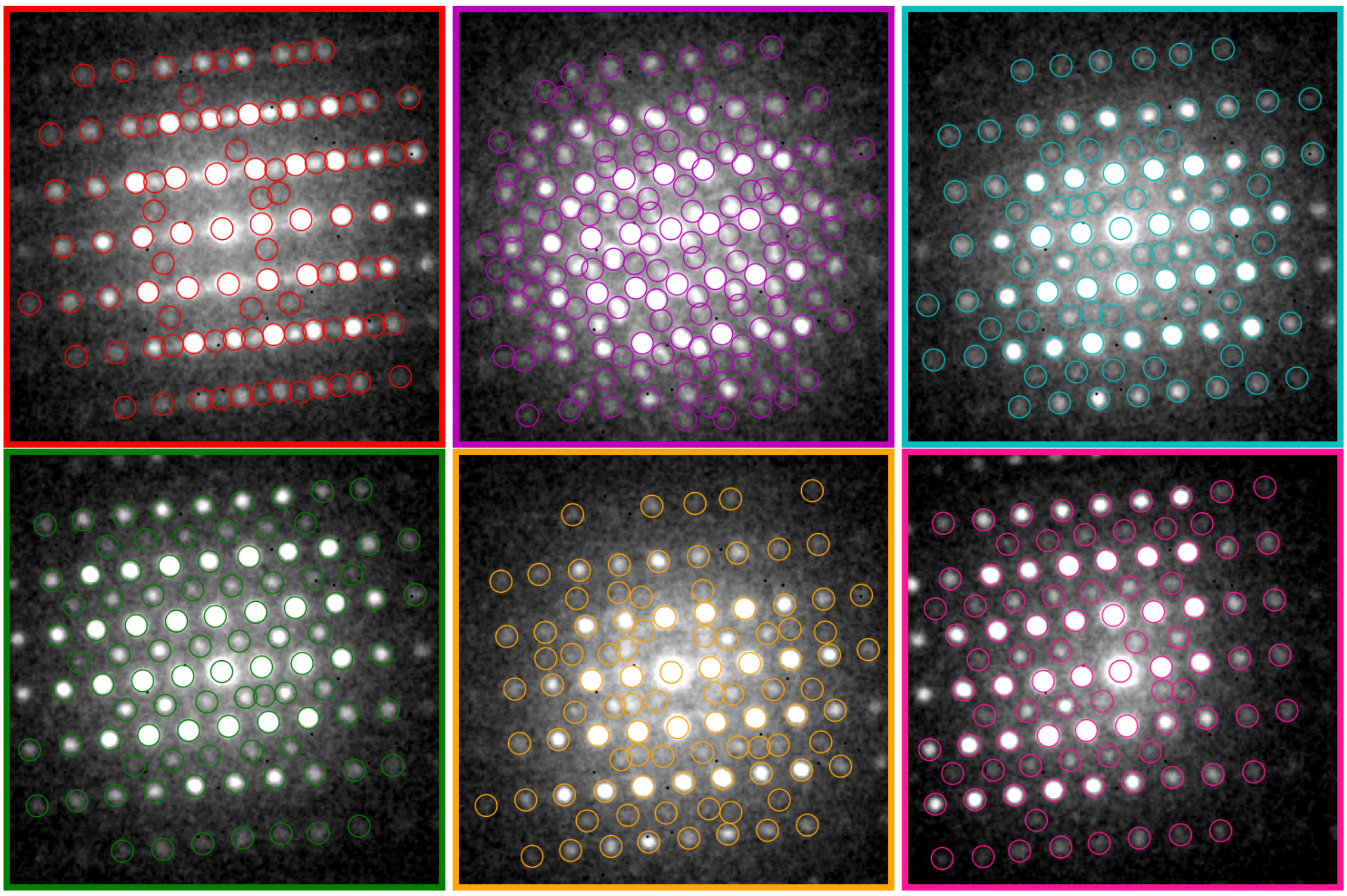

*Figure S3: Six different diffraction patterns extracted from the dataset to test the spot detection settings, with frames and spot circles colour-coded to the scan point colours in Figure S2.*